\documentclass[letterpaper]{article} 
\usepackage{aaai2026}  
\usepackage{times}  
\usepackage{helvet}  
\usepackage{courier}  
\usepackage[hyphens]{url}  
\usepackage{graphicx} 
\usepackage{natbib}  
\usepackage{caption} 
\usepackage{algorithm}
\usepackage{algorithmic}

\usepackage{multirow}
\usepackage{booktabs}
\usepackage{amsmath}
\usepackage{graphicx}
\usepackage{subcaption} 
\usepackage{makecell}
\usepackage{color}
\usepackage{amssymb}
\usepackage[switch]{lineno}
\usepackage[most]{tcolorbox}
\usepackage{xcolor}      

\usepackage{newfloat}
\usepackage{listings}
\DeclareCaptionStyle{ruled}{labelfont=normalfont,labelsep=colon,strut=off} 
\floatstyle{ruled}
\newfloat{listing}{tb}{lst}{}
\floatname{listing}{Listing}
\nocopyright 

\title{MHier-RAG: Multi-Modal RAG for Visual-Rich Document Question-Answering via Hierarchical and Multi-Granularity Reasoning}
\author{
    Ziyu Gong\textsuperscript{\rm 1},
    Chengcheng Mai\textsuperscript{\rm 2},
    Yihua Huang\textsuperscript{\rm 1}
}
\affiliations{
    \textsuperscript{\rm 1}State Key Laboratory for Novel Software Technology, School of Computer Science, Nanjing University\\
    \textsuperscript{\rm 2}School of Computer Science and Electronic Informatics, Nanjing Normal University\\

    ziyugong@smail.nju.edu.cn, yhuang@nju.edu.cn, maicc@njnu.edu.cn
}

\usepackage{bibentry}

\begin{document}

\maketitle

\begin{abstract}
The multi-modal long-context document question-answering task aims to locate and integrate multi-modal evidences (such as texts, tables, charts, images, and layouts) distributed across multiple pages, for question understanding and answer generation. The existing methods can be categorized into Large Vision-Language Model (LVLM)-based and Retrieval-Augmented Generation (RAG)-based methods. However, the former were susceptible to hallucinations, while the latter struggled for inter-modal disconnection and cross-page fragmentation. To address these challenges, a novel multi-modal RAG model, named MHier-RAG, was proposed, leveraging both textual and visual information across long-range pages to facilitate accurate question answering for visual-rich documents. A hierarchical indexing method with the integration of flattened in-page chunks and topological cross-page chunks was designed to jointly establish in-page multi-modal associations and long-distance cross-page dependencies. By means of joint similarity evaluation and large language model (LLM)-based re-ranking, a multi-granularity semantic retrieval method, including the page-level parent page retrieval and document-level summary retrieval, was proposed to foster multi-modal evidence connection and long-distance evidence integration and reasoning. Experimental results performed on public datasets, MMLongBench-Doc and LongDocURL, demonstrated the superiority of our MHier-RAG method in understanding and answering modality-rich and multi-page documents. 
\end{abstract}


\section{Introduction}
Document question-answering (Doc-QA) aims to answer questions based on document content. With the rise of Large Vision-Language Models (LVLMs), Doc-QA research \cite{V-Doc, OCR-VQA, VISA, SlideVQA, LayoutLLM} has transitioned from simple text-based approaches to more sophisticated multi-modal methods. In multi-modal long-context Doc-QA \cite{VisDoM, M3docrag, MDocAgent}, commonly seen in scientific papers, business reports, and instructional manuals, etc., relevant evidences needed to answer a question are often scattered across multiple pages and modalities, including texts, tables, charts, layouts and images, which poses challenges for complex multi-modal comprehension and long-distance reasoning over scattered content.

Existing methods for multi-modal long-context Doc-QA can be categorized into two types: LVLM-based methods \cite{Deepseek-vl, Internlm, Mplug-docowl} and RAG-based methods \cite{Re-Align, RULE, MMed-RAG}. LVLM-based methods directly processed the entire document by using large multi-modal models, jointly encoding both textual and visual features. However, these models tended to suffer context length limitations, insufficient long-range reasoning capabilities, and fact hallucinations. RAG-based methods focused on retrieving relevant evidences from the document before answer generation, which improved scalability but still failed to capture complex multi-modal dependencies and overlook cross-page evidence.

\begin{figure}
    \centering
    \begin{subfigure}{\linewidth}
        \centering
        \includegraphics[width=\linewidth]{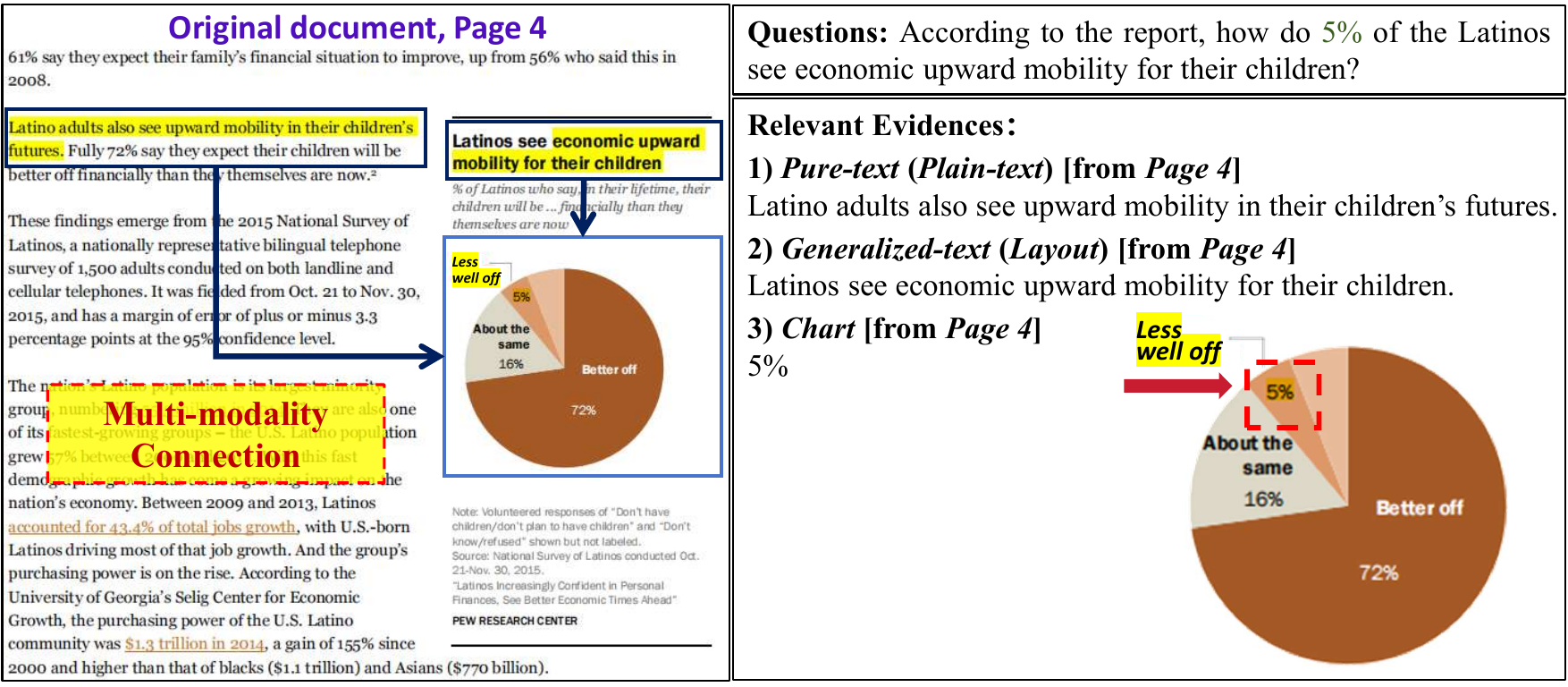}
        \caption{Multi-modal connection}
        \label{fig:Intro_subA}
    \end{subfigure}
    \begin{subfigure}{\linewidth}
        \centering
        \includegraphics[width=\linewidth]{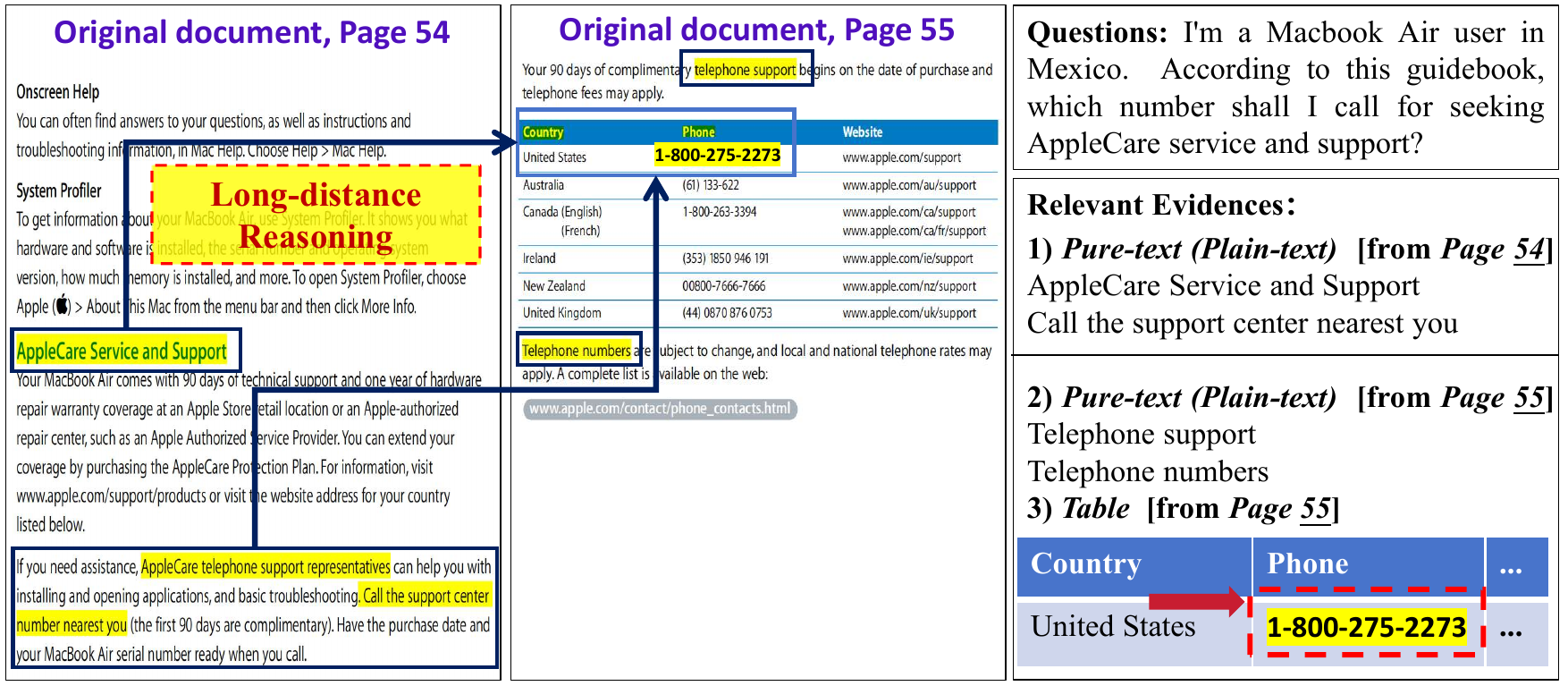}
        \caption{Long-distance reasoning}
        \label{fig:Intro_subB}
    \end{subfigure}
    \caption{Two challenges for multi-modal long-context document question-answering.}
    \label{fig:total}
\end{figure}

Based on the above research status of existing methods for the multi-modal Doc-QA, we listed two key challenges and outlined the corresponding solutions, as follows:

\textbf{Challenge 1: Absence of multi-modality connections.}
Multi-modal long-context document question-answering requires the integration of multi-modal information to synthesize accurate answers, rather than relying solely on either textual or visual cues. As shown in Figure \ref{fig:Intro_subA}, for the question ``how do 5\% of the Latinos see economic upward mobility for their children?'', several textual evidences are relatively easy to be retrieved, such as plain text and layout-based image caption, since the question shares common keywords and semantic meaning with these textual content. However, the actual answer ``Less well off'' resides within the visual image, which struggles to yield a high retrieval score due to few direct textual overlaps and semantic cues belonging to the pie chart. Therefore, it is essential to establish connections between the corresponding visual elements and surrounding textual information.

\textbf{Solution 1: Build multi-modal connections by combining in-page indexing structure with page-level parent page retrieval method.}
To address the challenge of multi-modality disconnection, a flattened in-page indexing strategy combined with the page-level parent-page retrieval was presented. Since semantically relevant textual and visual elements tend to co-occur in the same page, the parent page contains multi-modal retrieved contents related to the answer, which helps to bridge different modalities. This method transformed textual evidence into entry points for accessing associated visual information, enabling the model to aggregate multi-model evidence more effectively. 

\textbf{Challenge 2: Lack of cross-page evidence linking and reasoning ability.}
Another difficulty in multi-modal long-context Doc-QA is the isolation of evidence dispersed across different pages, which requires models to reason and aggregate long-distance cross-page evidences. As shown in Figure \ref{fig:Intro_subB}, for the question ``I'm a Macbook Air user in Mexico. According to this guidebook, which number shall I call for seeking AppleCare service and support?'', the relevant evidences span multiple modalities, including plain text and tables, and are distributed across different pages. Specifically, the actual phone number ``1-800-275-2273'' appears in a table on page 55, while the explanatory instruction, indicating that users should ``call the support center number nearest you'' is located on page 54. This demands the model to associate cross-page evidences and perform multi-step reasoning across multiple document sections, which is also not available in most existing models.

\textbf{Solution 2: Achieve long-distance reasoning by combining cross-page indexing structure with document-level summary retrieval method.}
To solve this problem, a topological cross-page indexing strategy combined with document-level summary-based retrieval was proposed. Semantically related content from different pages are grouped together through clustering and summarized by large language models, thereby promoting the retrieval scope to span across multiple pages. This method helped to aggregate evidence scattered across different pages, promoting long-distance reasoning.

Considering the above challenges and solution ideas, a novel retrieval-augmented generation method, named MHier-RAG, was proposed for the multi-modal long-context document question-answering task. The major contributions can be summarized as follows: 
\begin{itemize}
\item In proposed MHier-RAG, a hierarchical index structure with flattened in-page and topological cross-page chunks was proposed to establish correlations between diverse modalities and multiple pages.
\item A multi-granularity retrieval method with page-level parent page retrieval and document-level summary retrieval was also proposed to facilitate multi-modal evidence connection and long-distance evidence integration and reasoning. 
\item Extensive experiments conducted on two public datasets, MMLongBench-Doc and LongDocURL, verified the superiority and effectiveness of our method for the multi-modal long-context Doc-QA task.
\end{itemize}

\begin{figure*}[ht]
  \centering
  \includegraphics[width=\linewidth]{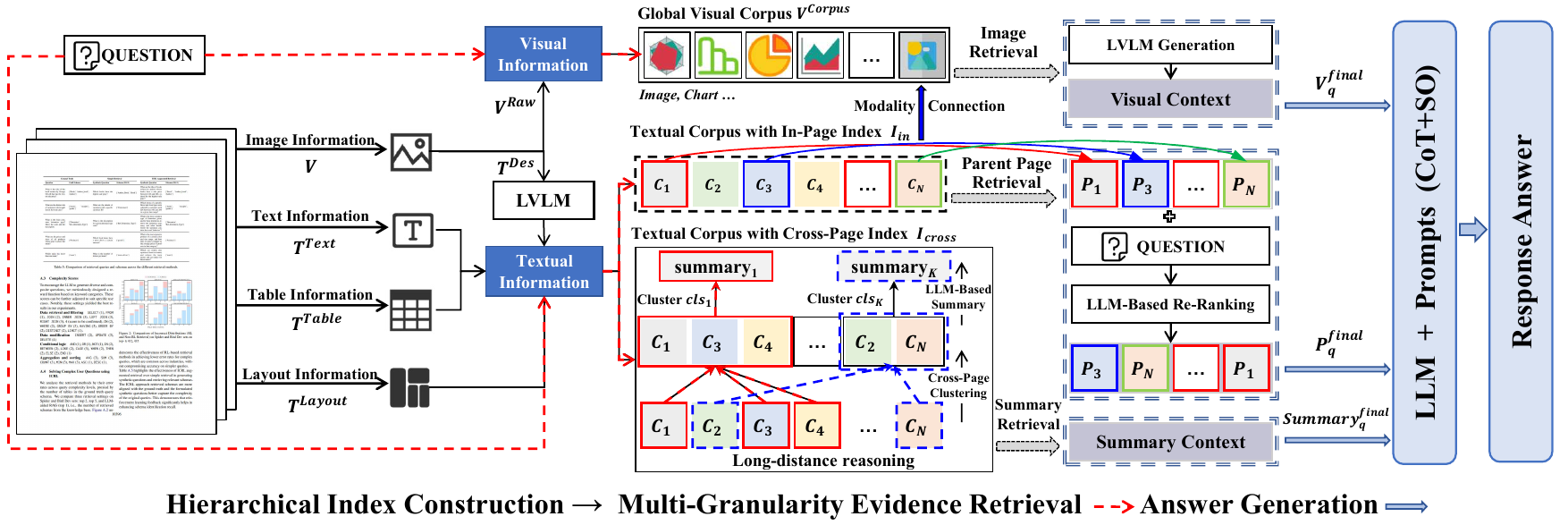}
  \caption{Overview of MHier-RAG with hierarchical index and multi-granularity retrieval for multi-modal Doc-QA. ($P_n$ is the parent page of $c_n$, $cls_K$ is the clustered block, `CoT' denotes chain-of-thought and `SO' denotes a structured output format.)}
  \label{fig:Overview}
\end{figure*}
\section{Our MHier-RAG Methodology}
\subsection{Task Description}
The objective of multi-modal long-context Doc-QA is to answer questions based on document content that contains both textual and visual content (e.g., texts, tables, charts, layouts and images), where relevant evidences may be dispersed across multiple pages and modalities. 

Figure \ref{fig:Overview} presents an overview of our MHier-RAG method. Given a query $Q$ and its corresponding document $D=\{d_1, d_2, ..., d_N\}$, where $N$ is denoted as the page number and $d_i$ represents the $i$-th page.The multi-modal long-context Doc-QA can be formalized as follows:
\begin{align}
    Y_{ans} &= {LLM}_{\theta}(y_i|y_{<i}, Q, R_q^{context}, P_{CoT}) \\
    R_q^{context} &= MMRetriever(HierIndex(D), Q)
\end{align}
where $Y_{ans}$ is the answer response generated by ${LLM}_{\theta}$, $R_q^{context}$ represents multi-modal retrieved evidences, $HierIndex(\cdot)$ is the hierarchical index structure for constructing retrievable corpus of $D$, $MMRetriever(\cdot)$ is the proposed multi-granularity retriever that searches for evidences related to $Q$ from document corpus, and $P_{CoT}$ is a Chain-of-Thought (CoT) prompting strategy, which is specifically designed to guide the model in generating step-by-step reasoning paths during answer generation.

\subsection{Hierarchical Index Construction Method with Multi-Modal Semantic Encoding}
For each page $d_i = \{T_i, V_i\}$ in document $D$, the textual content, $T_i$, consisted of three key components: (1) pure-text that preserved in original form; (2) tables that serialized into arranged sequences to retain its row-column-cell structure; (3) layout-based text. The composition of $T_i$ was denoted as $T_i = T_i^{Text} \cup T_i^{Table} \cup T_i^{Layout}$.

The visual context, $V_i$, was transformed into descriptive textual information via LVLMs, denoted as $T_i^{Des}$, converting visual semantics into text-based semantic encoding and retrieval space. Meanwhile, the raw images and charts have also been retained as the original and complete visual features, denoted as $V_i^{raw}$, to alleviate information loss. $V_i$ can be represented as $V_i = T_i^{Des} \cup V_i^{Raw}$.

Therefore, textual and visual content of documents can be separately defined as follows:
\begin{align}
    T^{corpus} &= T^{Text} \cup T^{Table} \cup T^{Layout} \cup T^{Des}\\
    V^{corpus} &= V^{Raw}
\end{align}

Based on the multi-modal content obtained above, two complementary hierarchical index structures were constructed at different levels, with the following formula:
\begin{equation}
    I = HierIndex(T^{Corpus}) = \{I_{in}, I_{cross}\}
\end{equation}
where $\boldsymbol{I_{in}}$ represents the flattened in-page index for establishing the association of different modality information within one page, and $\boldsymbol{I_{cross}}$ denotes the topological cross-page index for establishing the interaction of long-distance cross-page information.

\textbf{A. The Flattened In-Page Index.}
The flattened in-page indexing converted textual information (i.e., $T_i^{corpus}$ for each document page $d_i$) into a page-level list of smaller and uniformly-sized textual chunks, enabling direct access to pieces of evidence and facilitating finer-grained evidence extraction for retrieval. These textual chunks in $i$-th page of document $D$ can be denoted as $c_i = \{c_{i,1}, c_{i,2}, ..., c_{i,K_i}\}$, where $K_i$ is the chunk number of $i$-th page. Language models (LMs) were utilized for encoding the text attributes, thereby learning representations that capture their semantic meaning, denoted as $Z_i^c = LM(c_i)$. Thus, the flattened in-page index of document $D$ was defined as:
\begin{equation}
    I_{in} = \bigcup_{i=1}^N I(Z_i^c)
\end{equation}
where $N$ is the page number of document $D$ and $I(\cdot)$ represents the index of the encoded text chunks.

\textbf{B. The Topological Cross-Page Index.}
The topological cross-page indexing was conducted at the document-level scope, partitioning the entire textual content (i.e., $T^{corpus}$) into clustered blocks with similar semantic meanings for encoding and summarizing the similar semantics. These textual blocks (denoted as $B=\{b_1, b_2, ..., b_K\}$, where $K$ is the number of textual blocks in document $D$) were also encoded by language models, represented as $Z_i^b = LM(b_i)$.

Through iterative processing, all textual blocks were organized into a topological tree, where leaf nodes retained original attributes, intermediate nodes aggregated semantically related cross-page blocks with the gaussian mixture model for clustering the text blocks, and the root node summarized the topic-level semantics of the document with the large language model, thereby creating a multi-scale representation that captures both local multi-modal details and global cross-page document structure.

For instance, at each layer $l$, leaf embeddings $Z_i^{b,l-1}$ were clustered, and summaries were generated as follows:
\begin{align}
    G_{K_l}^{(l)} &= GMM_{Cluster}(\{Z_i^{b,l-1}\}) \\
    S_{K_l}^{(l)} &= \bigcup_{g_j^{(l)}\in G_{K_l}^{(l)}} LLM_{Summary}(g_j^{(l)})
\end{align}
where $K_l$ is the number of clusters, and then summaries were re-embed to form higher-layer nodes, denoted as $Z_i^{b,l} = LM(S_{K_l}^{(l)})$.

Therefore, the topological cross-page index of document $D$ was formulated as:
\begin{equation}
    I_{cross} = \bigcup_{l=0}^L \{I(Z_i^{b,l}) | i=1,2,...,K_l\}
\end{equation}
where $I(\cdot)$ represents the index of the encoded textual nodes.

\subsection{Multi-granularity Retrieval Method}
To alleviate the disconnection problem between multi-modal information and the difficulty of long-distance reasoning, a multi granularity content retriever with page-level parent-page retrieval by in-page index and document-level summary retrieval by cross-page index was presented, by searching for evidence related to the question in the corpus with hierarchical indexing. Our multi-granularity content retriever can be defined as follows: 
\begin{align}
& MMRetriever(HierIndex(D),Q) = \\
& MMRetriever(I_{in},Q) \cup MMRetriever(I_{cross},Q) \notag
\end{align}

\textbf{A. Page-level Parent Page Retrieval Method with LLM-based Re-ranking for Modality Connection.}
Since semantically related textual and visual evidence tend to be distributed on the same page, the parent-page retrieval method was proposed to augment the integration of multi-modal information. Firstly, we retrieved the textual content chunks most relevant to the question, and then associated chunks with their corresponding parent page. These parent pages contained more semantically similar visual and structural information, such as tables, layouts, charts, and images.

Specifically, given the query $Q$, we calculated the similarity between the embedded query and textual chunks with the flattened index $I_{in}$, and then selected the Top \textit{K} most relevant chunks based on semantic similarity as follows:
\begin{small}
\begin{equation}
C_q = \{c_{i,j}^c\} = arg TopK_{Z_{i,j}^c \in I_{in}} Sim(LM(Q),Z_{i,j}^c)
\end{equation}
\end{small} 
where $c_{i,j}^c$ is the $j$-th chunk on page $i$ of document $D$ and can be navigated to its source page $P_i$. The set of retrieved parent pages was obtained as $P_q = \{ParentPage(c_{i,j}^c|c_{i,j}^c \in C_q)\}$.

To further improve parent page retrieval, a fine-grained LLM-based re-ranking method was proposed to select pages that were more relevant to the question, where the large language model (LLM) was guided to rate the relevance between those retrieved pages $P_q$ and the problem $Q$ on a scale of 0 to 1. The final reordered pages based on the scores allocated by the LLM can be defined as:
\begin{equation}
P_q^{final} = arg TopK_{P_i \in P_q} LLM\_Score(Q,P_i)
\end{equation}

Meanwhile, we searched for images belonging to the parent page set $P_q^{final}$ in visual corpus $V^{Corpus}$ (denoted as $V_q$), and then used the Large Vision-Language Model (LVLM) to provide highly-relevant evidence related to the problem for the images (denoted as $V_q^{final}$), which can be formulated as follows:
\begin{align}
&V_q = \{ v_k \in P_q^{final} | v_k \in V^{Corpus} \} \\
&V_q^{final} = LVLM(Q,V_q)
\end{align}

Ultimately, through parent page retrieval and LLM-based re-ranking, the page-level multi-modal content retrieved by in-page indexing was defined as:
\begin{align}
    MMRetriever(I_{in},Q) = P_q^{final} + V_q^{final}
\end{align}

\textbf{B. Document-level Summary Retrieval Method for Long-Distance Reasoning.}
To associate and reason the long-distance evidence fragments across multiple pages, we utilized the topological indexing to achieve document-level retrieval of summaries across multiple pages, serving as a supplement to the page-level retrieval method.

Based on the given question $Q$, we calculated the semantic similarity between all nodes in the topological structure and the question, and then selected the Top \textit{K} relevant ones. Therefore, the document-level multi-modal content retrieved by cross-page indexing was denoted as follows:
\begin{align}
    &MMRetriever(I_{cross},Q) = Summary_q^{final}\\
    &= arg TopK_{Z_i^{b,l} \in I_{cross}} Sim(LM(Q), Z_i^{b,l}) \notag
\end{align}

\begin{table*}[t]
\centering
\small
\begin{tabular}{lcccccccccc}
\toprule
\multirow{2}{*}{\textbf{Model}} & \multicolumn{5}{c}{\textbf{Evidence Source}} & \multicolumn{3}{c}{\textbf{Evidence Page}} & \multirow{2}{*}{\textbf{Acc.}} & \multirow{2}{*}{\textbf{F1}} \\
\cmidrule(lr){2-6} \cmidrule(lr){7-9}
 &  TXT & LAY & CHA & TAB & FIG & SIN & MUL & UNA & &  \\
\midrule
\multicolumn{10}{l}{\textbf{$\textit{OCR(Tesseract\cite{Tesseract})}+\textit{Large Language Models(LLMs)}$}} \\
\hline
QWen-Plus \cite{Qwen-Plus} & 17.4 & 15.6 & 7.4 & 7.9 & 8.8 & 14.2 & 10.6 & 42.2 & 18.9 & 13.4\\
DeepSeek-V2 \cite{Deepseek-v2} & 27.8 & 19.6 & 8.8 & 17.0 & 9.4 & 20.2 & 15.4 & 48.1 & 24.9 & 19.6\\
Claude-3 Opus \cite{Claude} & 30.8 & \underline{30.1} & 16.4 & 24.4 & 16.3 & 32.0 & 18.6 & 30.9 & 26.9 & 24.5\\
Gemini-1.5-Pro \cite{Gemini-1.5} & 29.3 & 15.9 & 12.5 & 17.7 & 11.5 & 21.2 & 16.4 & \underline{73.4} & 31.2 & 24.8\\
GPT-4o \cite{GPT-4o} & \underline{41.1} & 23.4 & \underline{28.5} & \underline{38.1} & 22.4 & 35.4 & \underline{29.3} & 18.6 & 30.1 & 30.5 \\
\midrule
\multicolumn{10}{l}{\textbf{$\textit{Large Visual Language Models(LVLMs)}$}} \\
\hline
DeepSeek-VL \cite{Deepseek-vl} & 7.2 & 6.5 & 1.6 & 5.2 & 7.6 & 5.2 & 7.0 & 12.8 & 7.4 & 5.4\\
InternLM-XC2-4KHD \cite{Internlm} & 9.9 & 14.3 & 7.7 & 6.3 & 13.0 & 12.6 & 7.6 & 9.6 & 10.3 & 9.8\\
mPLUG-DocOwl 1.5 \cite{Mplug-docowl} & 8.2 & 8.4 & 2.0 & 3.4 & 9.9 & 7.4 & 6.4 & 6.2 & 6.9 & 6.3\\
Qwen-VL \cite{Qwen-vl} & 5.5 & 9.0 & 5.4 & 2.2 & 6.9 & 5.2 & 7.1 & 6.2 & 6.1 & 5.4\\
GPT-4V \cite{Gpt-4} & 34.4 & 28.3 & 28.2 & 32.4 & \underline{26.8} & \underline{36.4} & 27.0 & 31.2 & \underline{32.4} & \underline{31.2}\\
\midrule
\multicolumn{10}{l}{\textbf{$\textit{RAG methods}$}} \\
\hline
ColBERTv2 \cite{ColBERTv2} + LLaMA-3.1-8B & 23.7 & 17.7 & 14.9 & 24.0 & 11.9 & 25.7 & 12.2 & 38.1 & 23.5 & 19.7 \\
M3DocRAG \cite{M3docrag} (page=4) &  30.0 & 23.5 & 18.9 & 20.1 & 20.8 & 32.4 & 14.8 & 5.8 & 21.0 & 22.6 \\
\textbf{MHier-RAG (OURS)}(page=4)  & 41.6 & 30.2 & 40.9 & 48.9 & 25.1 & 48.5 & 31.7 & 74.9 & 48.2 & 41.4 \\
\textbf{MHier-RAG (OURS)}(page=10)  & \textbf{45.9} & \textbf{34.4} & \textbf{44.9} & \textbf{51.1} & \textbf{37.5} & \textbf{53.5} & \textbf{36.8} & \textbf{76.2} & \textbf{52.3} & \textbf{46.0} \\
\bottomrule
\end{tabular}
\caption{Experimental results on the MMLongBench-Doc dataset. The highest performance was bolded and the second best performance (except for ours) was underlined. `page=n' represents the number of pages retrieved from the parent page. `SIN', `MUL' and `UNA' separately denote singe-page, cross-page and unanswerable questions. }
\label{model_comparison}
\end{table*}

\begin{table}[t]
\centering
\small
\begin{tabular}{lc}
\toprule
\textbf{Model} &  \textbf{Acc.} \\
\midrule
\multicolumn{2}{l}{\textbf{$\textit{OCR(Tesseract)}+\textit{Large Language Models(LLMs)}$}} \\
\hline
LLaVA-OneVision \cite{LLaVA-OneVision}  & 23.3 \\
Qwen-VL \cite{Qwen-vl} & 25.0 \\
Gemini-1.5-Pro \cite{Gemini-1.5} & 32.0 \\
GPT-4o \cite{GPT-4o}  & 34.7 \\
O1-preview \cite{GPT-4o}  & 35.8 \\
\midrule
\multicolumn{2}{l}{\textbf{$\textit{Large Visual Language Models(LVLMs)}$}} \\
\hline
InternLM-XC2.5 \cite{Internlm} & 2.4 \\
mPLUG-DocOwl2 \cite{Mplug-docowl} & 5.3 \\
Pixtral \cite{Pixtral} & 5.6 \\
Llama-3.2 (Meta 2024) & 9.2 \\
LLaVA-OneVision \cite{LLaVA-OneVision} & 22.0 \\
Qwen-VL \cite{Qwen-vl} & 30.6 \\
\midrule
\multicolumn{2}{l}{\textbf{$\textit{RAG methods}$}} \\
\hline
ColBERTv2+LLaMA-3.1-8B & 49.1 \\
M3DocRAG \cite{SimpleDoc}(page=10) & \underline{52.2}\\
\textbf{MHier-RAG(OURS)}(page=4) & 52.4 \\
\textbf{MHier-RAG(OURS)}(page=10) & \textbf{57.2} \\
\bottomrule
\end{tabular}
\caption{Experimental results on the LongDocURL dataset. The highest performance was bolded and the second best performance (except for ours) was underlined.}
\label{model_comparison2}
\end{table}

\subsection{Answer Generation}
For each question $Q$, the retrieved evidences from corpus, such as $P_q^{final}$, $V_q^{final}$ and $Summary_q^{final}$, were integrated into the context window of the large language model (LLM), and the final answer was generated as follows:
\begin{align}
    Y_{ans} &= LLM_{\theta}(y_i|y_{<i}, Q, R_q^{context}, P_{CoT}) \\
    R_q^{context} &= P_q^{final} \cup V_q^{final} \cup Summary_q^{final}
\end{align}

To enhance response quality, a chain-of-thought (CoT) prompting method with a structured output (SO) format was also proposed for answer reasoning, which enabled direct extraction of final answers without the need for lengthy parsing. Details of prompts can be found in Appendix.

\section{Experiments}
\subsection{Dataset}
We evaluated models on two public datasets for multi-modal Doc-QA. \textbf{MMLongBench-Doc} \cite{MMLongBench} comprises 135 long PDF documents, each containing an average of 47.5 pages and 21,214 tokens. It consists of 1,082 expert-annotated questions. \textbf{LongDocURL} \cite{LongDocURL} is constructed upon 396 long PDF documents, with an average length  of 85.6 pages and 43,622.6 tokens. It collects 2,325 high-quality question-answer pairs. Their answers rely on evidences from multi-modalities and multi-pages.

\subsection{Implementation Details}
Docling \cite{Docling} was used for pdf parsing. The off-the-shelf LLMs/LVLMs were utilized for answer generation. All experiments were conducted on a single NVIDIA A100 GPU. More implementation details are in Appendix.

\subsection{Metrics} For MMLongBench-Doc and LongDocURL, we followed their official evaluation setups. We reported the accuracy of distinct evidence modality types and  evidence pages. The generalized accuracy and F1 score were also recorded.

\begin{table*}[t]
\small
\centering
\begin{tabular}{l|c|c|c|ccccccccc}
\toprule
\textbf{Variants} & \makecell[c]{\textbf{Parent Page} \\ \textbf{Retrieval}} & \makecell[c]{\textbf{Summary} \\ \textbf{Retrieval}} & \makecell[c]{\textbf{Visual} \\ \textbf{Info}} & \textbf{TXT} & \textbf{LAY} & \textbf{CHA} & \textbf{TAB} & \textbf{FIG} & \textbf{SIN} & \textbf{MUL} & \textbf{UNA} & \textbf{Acc.} \\
\midrule
\texttt{MHier-RAG} & \checkmark(page=10) & \checkmark & \checkmark & \textbf{45.9} & \textbf{34.4} & \textbf{44.9} & \textbf{51.1} & \textbf{37.5} & \textbf{53.5} & \textbf{36.8} & 76.2 & \textbf{52.3}  \\
\texttt{MHier-RAG\textsubscript{$v_1$}} & \checkmark(page=10) & \checkmark & $\times$ & 41.0 & 22.8 & 27.3 & 50.9 & 21.9 & 42.0 & 29.4 & 88.2 & 46.9  \\
\texttt{MHier-RAG\textsubscript{$v_2$}} & \checkmark(page=1) & \checkmark & \checkmark & 40.6 & 27.6 & 31.9 & 39.7 & 28.5 & 51.2 & 18.9 & 83.9 & 46.6\\
\texttt{MHier-RAG\textsubscript{$v_3$}} & \checkmark(page=1) & $\times$ & \checkmark & 34.7 &	25.1 & 28.6	& 34.3 & 25.7 &	47.2 & 13.9	& 84.8 & 43.3\\
\texttt{MHier-RAG\textsubscript{$v_4$}} & $\times$ & \checkmark & \checkmark & 26.4 & 15.6 &	22.8 & 23.8 & 21.9 & 32.2 & 14.0 & \textbf{90.1} & 37.5\\
\bottomrule
\end{tabular}
\caption{Ablation experiments for parent page retrieval with flattened in-page index, summary retrieval with topological cross-page index and visual information on the MMLongBench-Doc dataset.}
\label{ablation}
\end{table*}

\subsection{Main Results}
We compared our MHier-RAG with existing SOTA LVLM/LLM-based and RAG-based methods on the MMLongBench-Doc and LongDocURL datasets.

\textbf{MMLongBench-Doc.} Table \ref{model_comparison} listed the performance of models on the MMLongBench-Doc dataset. We observed that: (1) Both LVLM-based methods and LLM-based methods with OCR-parsed documents exhibited poor performance and struggled with multi-modal comprehension and long-distance reasoning for long-context document. Our MHier-RAG surpassed the best-performing LVLM, i.e., GPT-4V, by 19.9\% and 14.8\% in generalized accuracy and F1 score, respectively. (2) Compared with the RAG-based SOTA M3DocRAG, when the page number was four, MHier-RAG achieved superiority on all metrics, with improvements of 27.2\% and 18.8\% on generalized accuracy and F1 score. When the page number was set to ten, MHier-RAG achieved better performance with an accuracy of 52.3\% and a F1 score of 46.0\%. To be specific, the accuracy of visual charts (CHA) and figures (FIG) separately increased to 44.9\% and 37.5\%. The accuracy of multi-page (MUL) and unanswerable-questions (UNA) attained 36.8\% and 76.2\%, respectively. (3) These experimental results proved the advantages of our model in multiple modality understanding and long-distance reasoning, and the capability of alleviating hallucinations caused by LLMs.

\textbf{LongDocURL.} Table \ref{model_comparison2} showed the performance of models on the LongDocURL dataset. Similar phenomena were found: (1) Our MHier-RAG achieved better performance, compared to the top-performing LVLM (i.e., Qwen2-VL) and LLM (i.e., o1-preview). (2) MHier-RAG method surpassed the current SOTA M3DocRAG by a margin of 5\% in terms of generalized accuracy, when the page number was set to ten. (3) These results further validated the effectiveness of our approach for the connection of multi-modality and the link of cross-page evidence.

\begin{table}
\small
\centering
\begin{tabular}{l|c|c|c|c}
\toprule
\textbf{Variants} & \makecell[c]{\textbf{Parent Page} \\ \textbf{Retrieval}} & \makecell[c]{\textbf{Summary} \\ \textbf{Retrieval}}  & \makecell[c]{\textbf{Visual} \\ \textbf{Info}} & \textbf{Acc.} \\
\midrule
\texttt{MHier-RAG} & \checkmark(page=10) & \checkmark & \checkmark & \textbf{55.7}   \\
\texttt{MHier-RAG\textsubscript{$v_1$}} & \checkmark(page=10) & \checkmark & $\times$ & 53.1   \\
\texttt{MHier-RAG\textsubscript{$v_2$}} & \checkmark(page=1) & \checkmark & \checkmark & 50.4\\
\texttt{MHier-RAG\textsubscript{$v_3$}} & \checkmark(page=1) & $\times$ & \checkmark & 47.4\\
\texttt{MHier-RAG\textsubscript{$v_4$}} & $\times$ & \checkmark & \checkmark & 31.2\\
\bottomrule
\end{tabular}
\caption{Ablation experiments for parent page retrieval with flattened in-page index, summary retrieval with topological cross-page index and visual inforamtion on the LongDocURL dataset.}
\label{ablation2}
\end{table}

\subsection{Ablation Experiments}

\textbf{MMLongBench-Doc.}
Table \ref{ablation} presented ablation results on the MMLongBench-Doc dataset. After removing all visual information, the performance of the variant model \texttt{MHier-RAG\textsubscript{$_{v_1}$}} decreased to 46.9\%, which highlighted the importance of simultaneously integrating textual and visual information for the RAG-based method in Doc-QA. \texttt{MHier-RAG\textsubscript{$_{v_2}$}} restricted the number of parent page from ten to one, resulting in a performance decline to 46.6\%. The drop suggested that expanding the scope of parent page retrieval can enhance model performance. The removal of summary retrieval in \texttt{MHier-RAG\textsubscript{$_{v_3}$}} degraded the generalized accuracy to 43.3\%, indicating that the document-level summary retrieval is essential for aggregating evidences across multiple pages. After discarding the page-level parent page retrieval, \texttt{MHier-RAG\textsubscript{$_{v_4}$}} showed the lowest accuracy of 37.5\%, thereby confirming the importance of parent page retrieval in integrating multi-modal content.

It is worth noting that, the accuracy of unanswerable questions (UNA) in four variant models increased, suggesting that overly broad retrieval may introduce distractors in unanswerable cases. 

\textbf{LongDocURL.}
Table \ref{ablation2} reported similar ablation results on the LongDocURL dataset. The accuracy of \texttt{MHier-RAG\textsubscript{$_{v_1}$}} declined with the loss of visual information, proving the indispensability of multi-modal information. When the scope of parent page was restricted to one, as in \texttt{MHier-RAG\textsubscript{$_{v_2}$}}, the generalized accuracy dropped to 50.4\%, highlighting that retrieving a broader set of parent pages is important for contextual grounding oriented at question answering. The removal of summary retrieval in \texttt{MHier-RAG\textsubscript{$_{v_3}$}} reduced the accuracy to 47.4\%, emphasizing the critical role of topological cross-page summary chunks in capturing connections between dispersed evidence across pages. The worst-performing variant, \texttt{MHier-RAG\textsubscript{$_{v_4}$}}, which removed parent pages, yielded an average accuracy of 31.2\%, suggesting the role of parent page retrieval in enabling rich multi-modal comprehension. 

\begin{figure}
    \centering
    \begin{subfigure}{0.49\linewidth} 
        \includegraphics[width=\linewidth]{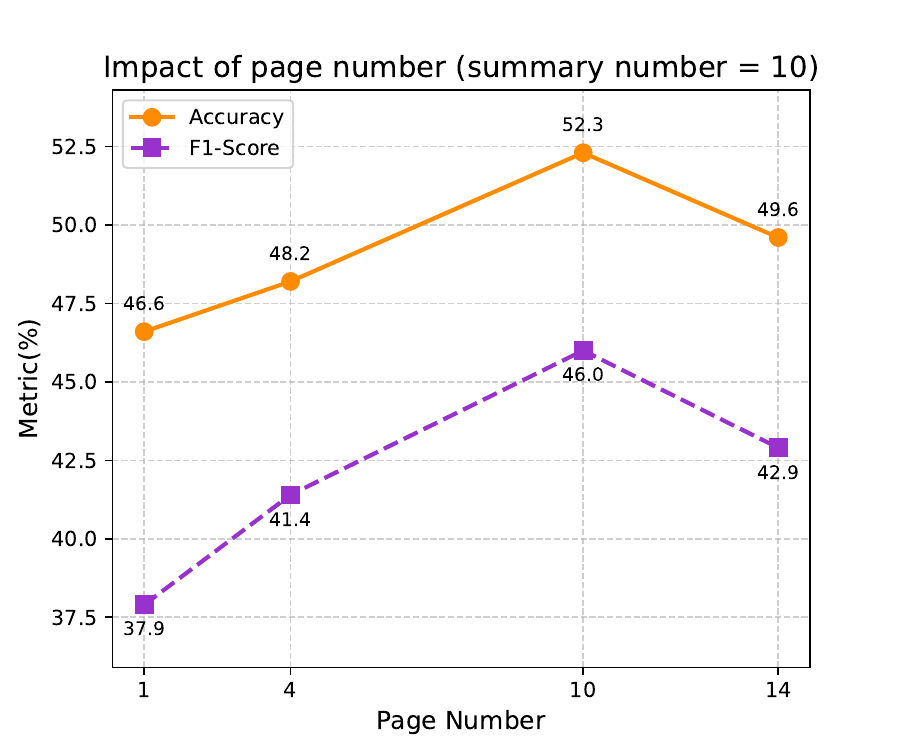}
        \caption{Page Number.}
        \label{fig:subA3}
    \end{subfigure}
    \begin{subfigure}{0.49\linewidth}
        \includegraphics[width=\linewidth]{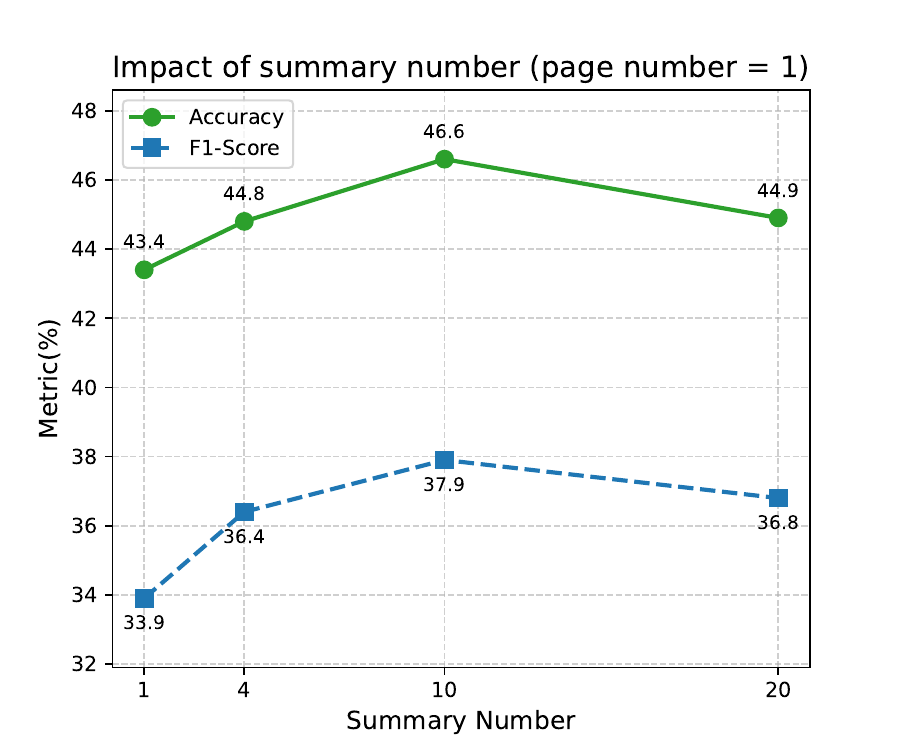}
        \caption{Summary Number.}
        \label{fig:subB3}
    \end{subfigure}
    \caption{The trend of our MHier-RAG model performance changing with the page number and summary number on the MMLongBench-Doc dataset.}
    \label{fig:Parameter}
\end{figure}

\subsection{Parameter Analysis on Content Size}
Figure \ref{fig:Parameter} showed the parameter analysis of retrieved page number and summary number for answer generation. MHier-RAG reached peak accuracy and F1 score when the parent page count was set to 10, but began to decline when the number of pages exceeded 10, which suggested that blindly increasing the page count may negatively impact performance. Meanwhile, as the number of summaries increased, the performance of MHier-RAG initially improved and subsequently declined, reaching its peak when the summary count was 10, which further supported the notion that an excessive amount of content may introduce irrelevant or distracting information, thereby negatively impacting the quality of the responses. Therefore, we set the number of pages and summaries to 10 to ensure that the retrieved content contains more evidence and avoids noisy information.

\begin{table}
\centering
\small
\begin{tabular}{l|c|c}
\toprule
\textbf{Cornerstone LLMs} & \makecell[c]{\textbf{MMLongBench-Doc} \\ \textbf{Acc.}} & \makecell[c]{\textbf{LongDocURL} \\ \textbf{Acc.}} \\
\midrule
\texttt{Qwen-turbo} & \textbf{52.3} & 55.7 \\
\texttt{GPT-4o} & 46.7 & \textbf{57.2} \\
\texttt{DeepSeek-chat} & 51.8 & 57.0 \\
\texttt{ERNIE-turbo} & 45.9 & 48.6 \\
\bottomrule
\end{tabular}
\caption{Extension on different LLMs for answer generation.}
\label{extension}
\end{table}

\begin{figure*}[t]
    \centering
    \begin{subfigure}{0.49\textwidth}  
        \includegraphics[width=\linewidth]{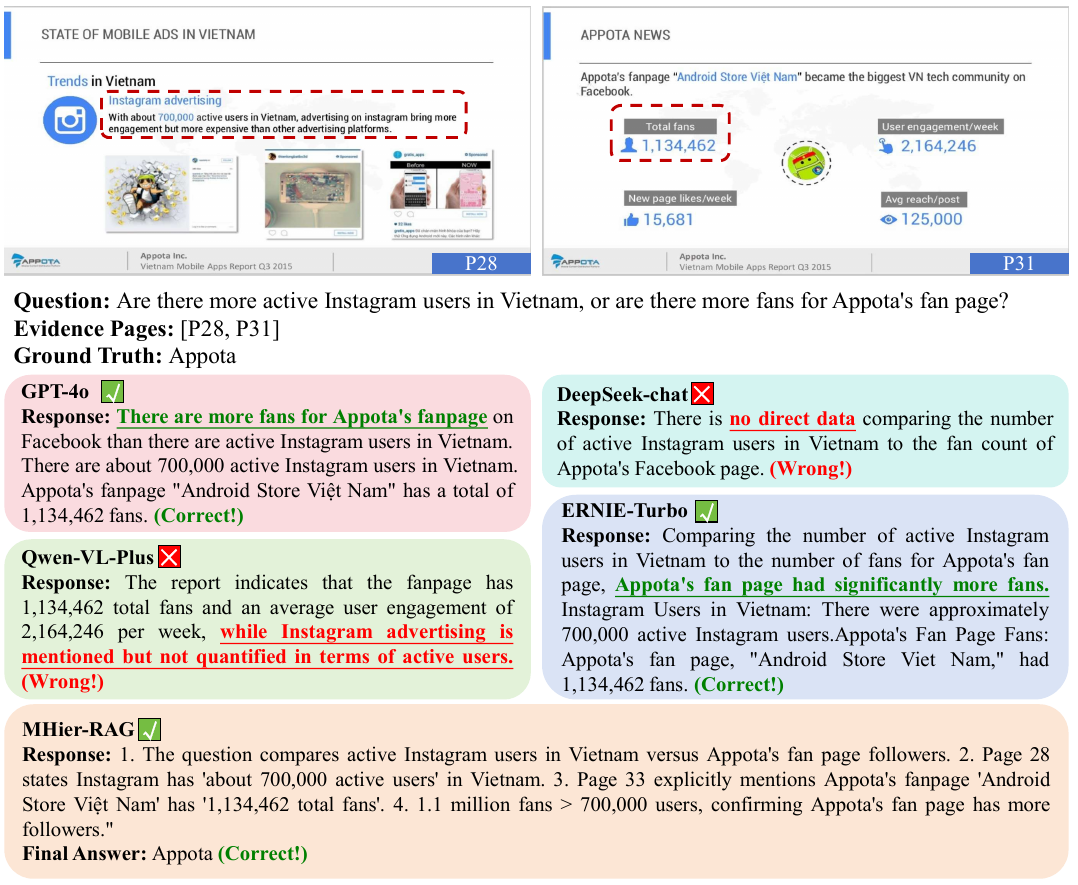}
        \caption{Test Case in MMLongBench-Doc.}
        \label{fig:Case1}
    \end{subfigure}
    \begin{subfigure}{0.49\textwidth}
        \includegraphics[width=\linewidth]{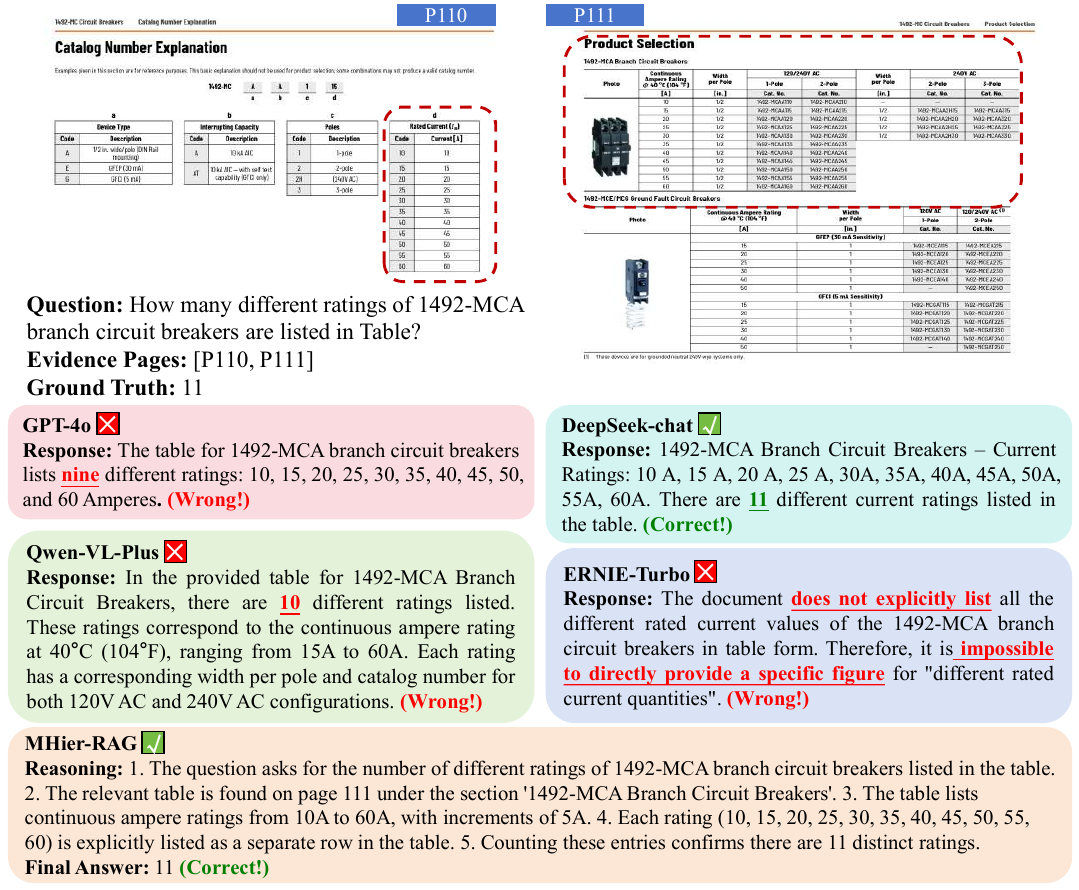}
        \caption{Test Case in LongDocURL.}
        \label{fig:Case2}
    \end{subfigure}
    \caption{Case Study on dataset MMLongBench-Doc and LongDocURL to compare the answer response of our MHier-RAG and LVLM-based methods (such as GPT-4o, DeepSeek-chat, Qwen-VL-Plus and ERNIE-Turbo).}
    \label{fig:Case_Study}
\end{figure*}

\subsection{Extension on LLMs for Answer Generation}
As shown in Table \ref{extension}, four different LLMs were selected as the cornerstone models to evaluate the performance impact on the document question-answering task. We found that Qwen-turbo achieved the highest accuracy at 52.3\% on the MMLongBench-Doc dataset, while GPT-4o attained the best performance with an accuracy of 57.2\% on the LongDocURL dataset. These results demonstrated that our proposed Retrieval-Augmented Generation (RAG)-based method, MHier-RAG, achieved excellent response on different large language models, which again validated the universality and transferability of our method.

\subsection{Case Study}
Figure \ref{fig:Case_Study} demonstrated two cases from the MMLongBench-Doc and LongDocURL datasets. The first question asked the comparison of fan numbers, with the answer required integrating textual statistics of Vietnam from page 28 with visual data of Appota on page 31. VLVM-based methods, such as Deepseek-chat and Qwen-VL-Plus, failed to answer the question. However, MHier-RAG correctly retrieved fan count from separate pages and concluded that Appota had more followers. The second question asked the number of 1492-MCA branch circuit breakers' rating, with the answer found in structured tables across page 110 and 111. GPT-4o and Qwen-VL-Plus gave incorrect answers of 9 and 10, respectively, which failed to identify all the entries. However, MHier-RAG correctly located the relevant table and identified the rating values. In summary, these cases again demonstrated that our method is adaptable for the comprehension of multi-modal and multi-page evidences that retrieved from document content. 

\section{Related Work}
Document question-answering has progressed from processing textual documents \cite{TriviaQA, Natural-Questions, TAT-DQA} to tackle lengthy documents involved with multi-modal elements and complicated structures across multiple pages \cite{MMLongBench, LongDocURL, UDA}, which demands capability of modality comprehension and long-distance reasoning. 

\textbf{LVLM-based Methods for Multi-modal Doc-QA Task.}
Large Vision-Language Models \cite{Qwen-vl, Mplug-docowl, Internlm} were regarded as an effective solution to handle multi-modal document question-answering, since they combined the deep linguistic capabilities of large language models with advanced visual processing for document images. However, Ma et al. \cite{MMLongBench} and Deng et al. \cite{LongDocURL} 
indicated that LVLMs still faced challenges in integrating evidences from different modalities and pages, and were prone to hallucinations.

\textbf{RAG-based Methods for Multi-modal Doc-QA Task.}
Traditional Retrieval-Augmented Generation (RAG)-based models \cite{Text-only, ColBERT} exhibited a uni-modality bias, predominantly relying on textual information while inadequately incorporating visual evidence from documents. To fully exploit visual elements within documents, Colpali \cite{ColPali}, DSE \cite{DSE} and VisRAG \cite{VisRAG} directly encoded the images of document pages for retrieval. MDocAgent \cite{MDocAgent} separately used text-based and image-based agents to handle textual and visual information, thereby obtaining critical information within their respective modalities in the retrieval phrase and generating refined answers. However, these existing RAG-based methods often overlooked the mutual connections between different modalities of information and remained inadequate in addressing cross-page integration and reasoning challenges. 

\section{Conclusion}
A retrieval-augmented generation method (MHier-RAG) was presented for multi-modal long-context document question-answering. A hierarchical index structure with flattened in-page and topological cross-page index was constructed to establish multi-modal connection and long-distance linkage. A multi-granularity retrieval with page-level parent page retrieval and document-level summary retrieval was proposed for searching required evidences, which were scattered across multi-modalities and multi-pages. Experiments conducted on two public datasets demonstrated the superiority of MHier-RAG in multi-modal long-context Doc-QA.


\section{Acknowledgments}
This work was supported by the National Natural Science Foundation of China (Nos.61572250 and 62476135). Jiangsu Province Science \& Tech Research Program(BE2021729), Open project of State Key Laboratory for Novel Software Technology, Nanjing University (KFKT2024B53), Jiangsu Province Frontier Technology Research and Development Program (BF2024005), Nanjing Science and Technology Research Project (202304016) and Collaborative Innovation Center of Novel Software Technology and Industrialization, Jiangsu, China.

\bibliography{aaai2026}

\newpage
\section{Appendix}
\subsection{Research Motivation}
Multi-modal long-context document question-answering (Doc-QA) involves answering queries by analyzing and integrating evidences across texts, tables, charts, images and layouts within multiple pages, requiring the capability of multi-modal connection and long-distance reasoning. However, there are currently few multi-modal long-context Doc-QA methods equipped with these abilities, which are worth further research. 

\begin{figure}[ht]
  \centering
  \includegraphics[width=\linewidth]{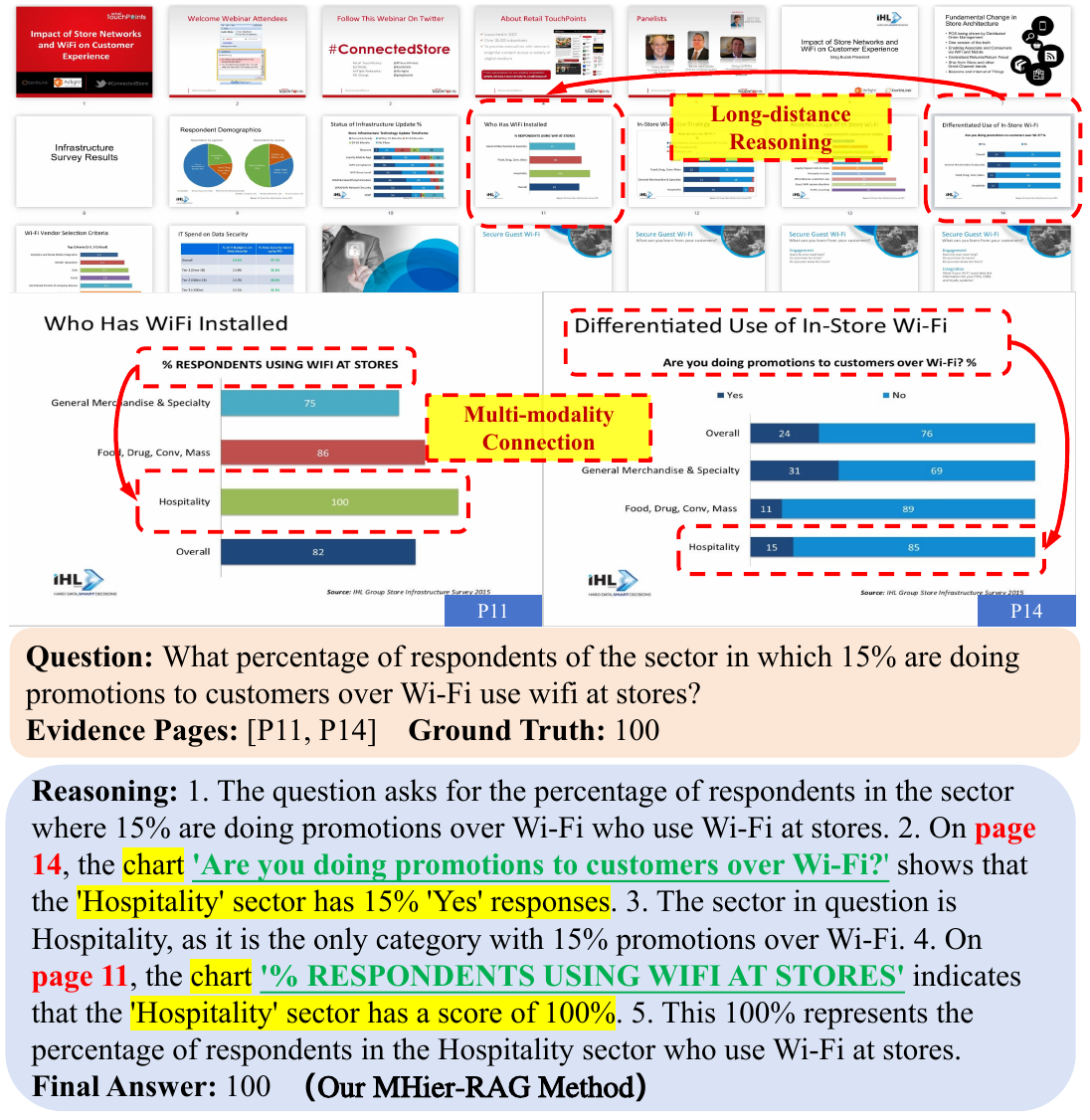}
  \caption{Necessities of multi-modal connection and long-distance reasoning for multi-modal long-context Doc-QA methods.}
  \label{fig:Overview2}
\end{figure}

Figure \ref{fig:Overview2} illustrates the capabilities required for multi-modal long-context Doc-QA and the advantage of our proposed MHier-RAG method. For the question ``What percentage of respondents of the sector in which 15\% are doing promotions to customers over Wi-Fi use wifi at stores?'', evidences required for the answer are scattered in the visual charts on page 11 and page 14. To correctly answer this question, multi-modal Doc-QA methods need to establish multi-modal connections. Due to the common keywords and similar semantics, the question is relatively easy to retrieve the textual titles ``\% RESPONDENTS USING WIFI AT STORES'' in page 11 and ``Are you doing promotions to customers over Wi-Fi?'' in page 14. Therefore, it is crucial to associate these textual captions with their surrounding related visual bar charts. Multi-modal Doc-QA methods also need to achieve long-distance reasoning to synthesize information on these non-consecutive pages. To be specific, Doc-QA methods first require to identify the ``Hospitality" industry (the only category with a 15\% promotion from the visual chart on page 14), and then retrieve its corresponding 100\% Wi-Fi usage rate from the data on page 11.

Our MHier-RAG first confirmed from page 14 that hospitality is the only industry that conducts a 15\% promotion through Wi-Fi, then found the proportion of respondents in the hotel industry who use Wi-Fi in stores from page 11, and finally concluded the correct answer 100 through reasonable multi-step reasoning. This indicates that our MHier-RAG has the ability of modality correlation and long-distance reasoning, which is crucial for integrating multi-modal cross-page evidences to answer questions.

\subsection{Implementation Details}
In the settings of our MHier-RAG, we used docling to achieve pdf parsing and used off-the-shelf Large Vision-Language Models (such as Qwen-vl-plus) to describe visual information in documents. For the flattened in-page index, the chunk size was set to 300 tokens (about 15 sentences) and encoded by text-embedding-v4 provided by Qwen. To avoid information loss due to text split, a text overlap with 50 tokens was added. For the topological cross-page index, the chunk size was set to 100 (about 5 sentences) and encoded by multi-qa-mpnet-base-cos-v1 provided by Sentence-Transformer. The gaussian mixture model was used for chunk clustering, and Gpt-3-turbo was adopted for summarizing the clustered chunks. For multi-granularity evidence retrieval, we selected 10 highest-scored parent pages and summaries as the input context for each question. We used Large Vision-Language Models (such as Qwen-turbo) to conduct retrieval re-ranking. We directly used off-the-shelf LLMs for answer generation. All experiments were conducted on a single NVIDIA A100 GPU.

\subsection{Prompt Setting}
For answer generation in multi-modal long-context document question-answering, a multi-step reasoning method was proposed for encompassing evidence curation and chain-of-thought reasoning for the retrieved relevant sources. 
\subsubsection{Prompt for Answer Generation}
The prompt for answer generation can be divided into several parts: 
\begin{itemize}
    \item Template for Context and Questions
    \item General Guidelines
    \item Response Formats
    \item Question-Answer Examples
\end{itemize} 

\textbf{A. Template for Context and Questions}
\begin{tcolorbox} 
Here is the context:\{context\} Here is the question:\{question\}
\end{tcolorbox}

\textbf{B. General Guidelines}

\begin{tcolorbox}
You are a RAG (Retrieval-Augmented Generation) answering system.
Your task is to answer the given question based only on information from the pdf report, which is uploaded in the format of relevant evidences extracted using RAG. \\

Before giving a final answer, carefully think out loud and step by step. Pay special attention to the wording of the question.\\
- Keep in mind that the content containing the answer may be worded differently than the question.\\
- The question was autogenerated from a template, so it may be meaningless or not applicable to the given report.
\end{tcolorbox}

\textbf{C. Response Formats}

The response format consists of four parts: (1) Step By Step Analysis, (2) Reasoning Summary, (3) Relevant Pages, and (4) Final Answer.

\begin{tcolorbox}[title = {Step By Step Analysis}] 
Detailed step-by-step analysis of the answer with at least 5 steps and at least 150 words. Pay special attention to the wording of the question to avoid being tricked.
\end{tcolorbox}

\begin{tcolorbox}[title = {Reasoning Summary}] 
Concise summary of the step-by-step reasoning process. Around 50 words.
\end{tcolorbox}

\begin{tcolorbox}[title = {Relevant Pages}] 
List of page numbers containing information directly used to answer the question. Include only: \\
- Pages with direct answers or explicit statements. \\
- Pages with key information that strongly supports the answer. \\
Do not include pages with only tangentially related information or weak connections to the answer. At least one page should be included in the list.
\end{tcolorbox}

\begin{tcolorbox}[title = {Final Answer}] 
Note: different prompts for different data types. 
\end{tcolorbox}

The expected answer data for questions includes various types, such as List, Integer, String and Float. Different prompt variants have been proposed to achieve higher-quality responses.

\begin{tcolorbox}[title = {Final Answer [List]}] 
A list of values extracted from the context. Each value should be: \\
- For strings: exactly as it appears in the context \\
- For numbers: converted to appropriate type (int or float) \\
- For nested lists: maintain the original structure \\
- Return `Not answerable' if information is not available in the context 
\end{tcolorbox}

\begin{tcolorbox}[title = {Final Answer [Integer]}] 
An integer value is expected as the answer. \\
- Pay attention to units (thousands, millions, etc.) and adjust accordingly \\
- Round to nearest integer if necessary \\
- Return `Not answerable' if: \par
 \quad - The value is not an integer \par
 \quad - Information is not available \par
 \quad - Currency mismatch occurs 
\end{tcolorbox}

\begin{tcolorbox}[title = {Final Answer [String]}] 
A string value is expected as the answer. \\
- Extract exactly as it appears in the context \\
- Do not modify or summarize the text \\
- Return `Not answerable' if information is not available in the context
\end{tcolorbox}

\begin{tcolorbox}[title = {Final Answer [Float]}] 
A floating-point number is expected as the answer. \\
- Maintain original decimal precision from the context \\
- Pay attention to units (thousands, millions, etc.) and adjust accordingly \\
- Return `Not answerable' if: \par
 \quad - The value is not a number \par
 \quad - Information is not available \par
 \quad - Currency mismatch occurs
\end{tcolorbox}

\textbf{D. Question-Answer Examples}

Depending on different data types, several question-answer examples are also provided to improve response quality, which contain step by step analysis, reasoning summary, relevant pages and final answer. It helps the large language model to respond required structured output format for different answer types.

Examples of answer type as a list, a float value, an integer value and a string are shown as follows:

\begin{tcolorbox}[title = {Examples [List]}] 
\textbf{Question} \\
What are the quarterly revenue figures for Apple Inc. in 2022?
\tcblower
\textbf{Answer} \\
\textbf{step\_by\_step\_analysis:} 1. The question asks for quarterly revenue figures for Apple Inc. in 2022, which implies we need to find four distinct values corresponding to each quarter. 2. Examining the context, we find a table titled `Quarterly Financial Results' on page 45 that lists revenue figures for each quarter of 2022. 3. The table shows: Q1(123.9B), Q2(97.3B), Q3(82.96B), Q4(90.15B). 4. We verify these are indeed revenue figures by checking the column header and accompanying notes. 5. The values are extracted exactly as presented, converted to float type for consistency. \\

\textbf{reasoning\_summary:} The `Quarterly Financial Results' table on page 45 provides the exact quarterly revenue figures for 2022, which are extracted and converted to float values. \\

\textbf{relevant\_pages:} [45] \\

\textbf{final\_answer:} [123.9, 97.3, 82.96, 90.15]
\end{tcolorbox}

\begin{tcolorbox}[title = {Examples [Integer]}] 
\textbf{Question} \\
How many employees did Tesla Inc have at the end of 2022?
\tcblower
\textbf{Answer} \\
\textbf{step\_by\_step\_analysis:} 1. The question asks for the number of Tesla Inc. employees at the end of 2022, which should be a whole number. 2. On page 56 of the annual report, we find the statement: `As of December 31, 2022, we employed approximately 127,855 full-time employees worldwide.' 3. The number 127,855 is explicitly stated as the employee count. 4. We verify this is a global total by checking the context which mentions `worldwide'. 5. No unit conversion is needed as this is already a direct count. \\

\textbf{reasoning\_summary:} Page 56 explicitly states Tesla employed 127,855 full-time employees worldwide as of December 31, 2022. \\

\textbf{relevant\_pages:} [56] \\

\textbf{final\_answer:} 127855
\end{tcolorbox}

\begin{tcolorbox}[title = {Examples [String]}] 
\textbf{Question} \\
What's the percentage of people who are democrats and voted in the last election compared to the entire population in 2024?
\tcblower
\textbf{Answer} \\
\textbf{step\_by\_step\_analysis:} 1. Question requires two precise data points: democrat voters and total population 2. Searched for `2024 election' references - none found 3. Checked all demographic sections - no voting breakdown by party 4. Verified document metadata - report finalized Q3 2023 (pre-election) 5. Attempted alternative queries - no matching tables/charts 6. Conclusion: Data unavailable in this report \\

\textbf{reasoning\_summary:} Document contains no 2024 election data (pre-dates election) and lacks democrat-specific voting percentages, making question unanswerable. \\

\textbf{relevant\_pages:} [] \\

\textbf{final\_answer:} Not Answerable
\end{tcolorbox}

\begin{tcolorbox}[title = {Examples [Float]}] 
\textbf{Question} \\
What was the gross profit margin percentage for NVIDIA Corporation in Q3 2022?
\tcblower
\textbf{Answer} \\
\textbf{step\_by\_step\_analysis:} 1. The question asks for NVIDIA's gross profit margin percentage in Q3 2022, which should be a decimal number. 2. On page 32 of the quarterly report, we find the statement: `Gross margin for the quarter was 53.6\%, down from 56.1\% in the prior quarter.' 3. The value 53.6\% is explicitly stated as the gross margin for the quarter. 4. We verify this is for Q3 2022 by checking the report header and date. 5. The percentage is converted to its decimal equivalent (53.6). \\

\textbf{reasoning\_summary:} Page 32 states NVIDIA's Q3 2022 gross margin was 53.6\%, which is converted to the decimal value 53.6. \\

\textbf{relevant\_pages:} [32] \\

\textbf{final\_answer:} 53.6
\end{tcolorbox}

\subsubsection{Prompt for LLM-Based Re-Ranking}
The prompt for LLM-based re-ranking in the page-level parent page retrieval can be devided into two parts: (1) Template for Pages and Questions, (2) General Guidelines.

\textbf{A. Template for Pages and Questions}

\begin{tcolorbox}
Here is the query: \{query\}  Here is the retrieved text block: \{retrieved\_page\} 
\end{tcolorbox}

\textbf{B. General Guidelines}

\begin{tcolorbox}
You are a RAG (Retrieval-Augmented Generation) retrievals ranker. You will receive a query and retrieved text block related to that query. Your task is to evaluate and score the block based on its relevance to the query provided. 
\tcblower
Instructions: 

\textbf{1. Reasoning:} \\
   Analyze the block by identifying key information and how it relates to the query. Consider whether the block provides direct answers, partial insights, or background context relevant to the query. Explain your reasoning in a few sentences, referencing specific elements of the block to justify your evaluation. Avoid assumptions—focus solely on the content provided. 

\textbf{2. Relevance Score (0 to 1, in increments of 0.1):} \\
   0 = Completely Irrelevant: The block has no connection or relation to the query. \\
   0.1 = Virtually Irrelevant: Only a very slight or vague connection to the query. \\
   0.2 = Very Slightly Relevant: Contains an extremely minimal or tangential connection. \\
   0.3 = Slightly Relevant: Addresses a very small aspect of the query but lacks substantive detail. \\
   0.4 = Somewhat Relevant: Contains partial information that is somewhat related but not comprehensive. \\
   0.5 = Moderately Relevant: Addresses the query but with limited or partial relevance. \\
   0.6 = Fairly Relevant: Provides relevant information, though lacking depth or specificity. \\
   0.7 = Relevant: Clearly relates to the query, offering substantive but not fully comprehensive information. \\
   0.8 = Very Relevant: Strongly relates to the query and provides significant information. \\
   0.9 = Highly Relevant: Almost completely answers the query with detailed and specific information. \\
   1 = Perfectly Relevant: Directly and comprehensively answers the query with all the necessary specific information. 

\textbf{3. Additional Guidance:} \\
   - Objectivity: Evaluate block based only on their content relative to the query. \\
   - Clarity: Be clear and concise in your justifications. \\
   - No assumptions: Do not infer information beyond what's explicitly stated in the block.
\end{tcolorbox}
\end{document}